# 3D ordered composites of nanopolyacetylene: Mechanism of self-organization and properties

Valerii Kobryanskii[1,2]

[1]*Institute of Chemical Physics RAS, Moscow 117977 Russia.* [2]*Supermat Intl. New York 10475 USA.*

**Here we demonstrate that the key pre-condition to the outstanding characteristics of nanopolyacetylene (NPA) is the 3D coherent ordering of nano-particles in polymer matrix. We have found that pumping out, or exposing of films in dry atmosphere leads to removal of water and a) disappearance of vibrational structure from the optical absorption spectra, b) appearance of background absorption in near-IR region , c) reduction of intensity of Raman scattering, and d) sharp decrease in stability under laser irradiation. The results afford a connection between optical properties of NPA and the three-dimensional coherent order arising from self-organization of nano-particles at synthesis and during film formation. The driving force behind self-organization is the abnormal high vibrational hyper-polarizability of macromolecules leading to coherent dispersive interactions between them. The critical condition for 3D coherent ordering is vibrational and rotational mobility of nano-particles, provided by plasticized layers of water on their surface. The principles of synthesis employed can further be used to create 3D coherently ordered photonic crystals and meta-materials.**

1. Introduction to 3D ordered photonic crystals and meta-materials.

Creation of 3D arranged nano-materials is one of the fundamental problems of material science. Theoretical models and experiments on existing nano-materials demonstrate that ensembles of 3D arranged nano-particles can possess exceptional optical and electronic properties not observed in bulk materials, or in nature [1]. The scientific community has allotted such ensembles into a separate class of photonic crystals and meta-materials [2,3] and is expending considerable efforts to elaborate them. For nano-materials, this typically involves methods of self organization proposed for colloid solutions [1], which allow obtaining 3D arranged nano-materials with large volume fraction of nano-particles [4]. However, development of 3D arranged nano-materials containing insignificant volume fraction of nano-particles remains in the category of problems well-stated, with no solutions proposed [1]. Such arrangement assumes interaction between nano-particles located at significant distances from each other, sufficient for their self-organization at thermal equilibrium. Existing models assume that Coulomb's forces drive self-organization of nano-particles in colloid solutions [5,6]. They explain the formation of colloid and supra-crystals, but assume that an ensemble of nano-particles located at significant distances from each other is thermodynamically unstable [1].

2. Introduction to 3D ordered conjugated polymers.

The pre-requisites to application of conjugated polymers (CP) in photonics and electronics are a) properties on the par with inorganic materials, and b) high stability at interaction with electromagnetic fields, and under optical irradiation [7,8]. Unfortunately, CP are either insoluble, or form a disordered solid phase from solutions, and their electronic structure is very sensitive to chemical and structural defects [9,10]. A path to applications is to transition from bulk materials to highly ordered nano-composites. In 1987-89s we synthesized soluble nano-composites of polyacetylene (NPA) [11,12] with an abnormally high intensity of Raman scattering [13], and high stability under laser radiation [14].

This article discusses the optical properties of 3-D ordered nano-composites of polyacetylene, and offers a model connecting 3-D self-organization of nano-particles with resonant interaction between them. Possible role of water as the enabling medium for such interaction is also discussed.

3. Experimental.

Compositions of NPA were obtained by polymerization of acetylene in the solution of polyvinylbutyral in n-butanol on the rhenium catalyst [12]. During polymerization the viscosity of reaction solution increases with formation of gel, which contains nano-particles of PA. Gel dissolves in the surplus of n-butanol during mixing and it easily forms films upon drying. The 5 - 500 μm thickness films, containing ~1% of nano-particles, were prepared from reaction mixture by slow evaporation of solvent at horizontally located supports. Process of film formation carried out in ventilated chamber within 24 hours at temperature 25$^o$C and relative humidity of 50 %. The equilibrium quantity of water in the films was determined from the difference of their weight before and after pumping out. At temperature 25$^o$C and relative humidity of 50 % the initial films contain ~1.4 % of water. Absorption spectra in UV-Vis-NIR field were measured in vacuum cuvettes at spectrophotometers HP-8453 and Perkin-Elmer "Lambda 9". Raman scattering spectra were measured in vacuum cuvettes at spectrometer Bruker-IFS 100 with a Nd-Yag 1064nm laser. FTNIR and FTIR spectra were measured with spectrometer Bruker-IFS28 at free standing films, which were placed into its chamber blown by nitrogen, with relative humidity <1 %.

## 4. Optical properties of NPA films.

The first set of figures presents UV-Vis (Fig.1a) and near-IR (Fig.1b) absorption spectra of NPA films, along with Raman scattering spectra (Fig.1c), as record on air (curve 1) and in vacuum 10$^{-3}$ mmHg (curve 2). The spectra differ sharply between the two environments: pumping out of films leads to deterioration of all spectral characteristics. In electronic absorption spectrum (Fig.1a), resolution of vibrational structure worsens. In near-IR spectrum (Fig.1b), long wave length tail appears, which is characteristic of a disordered CP. Raman scattering spectra (Fig.1c) exhibit a sharp reduction of intensity of fundamental vibrations of polyene chains.

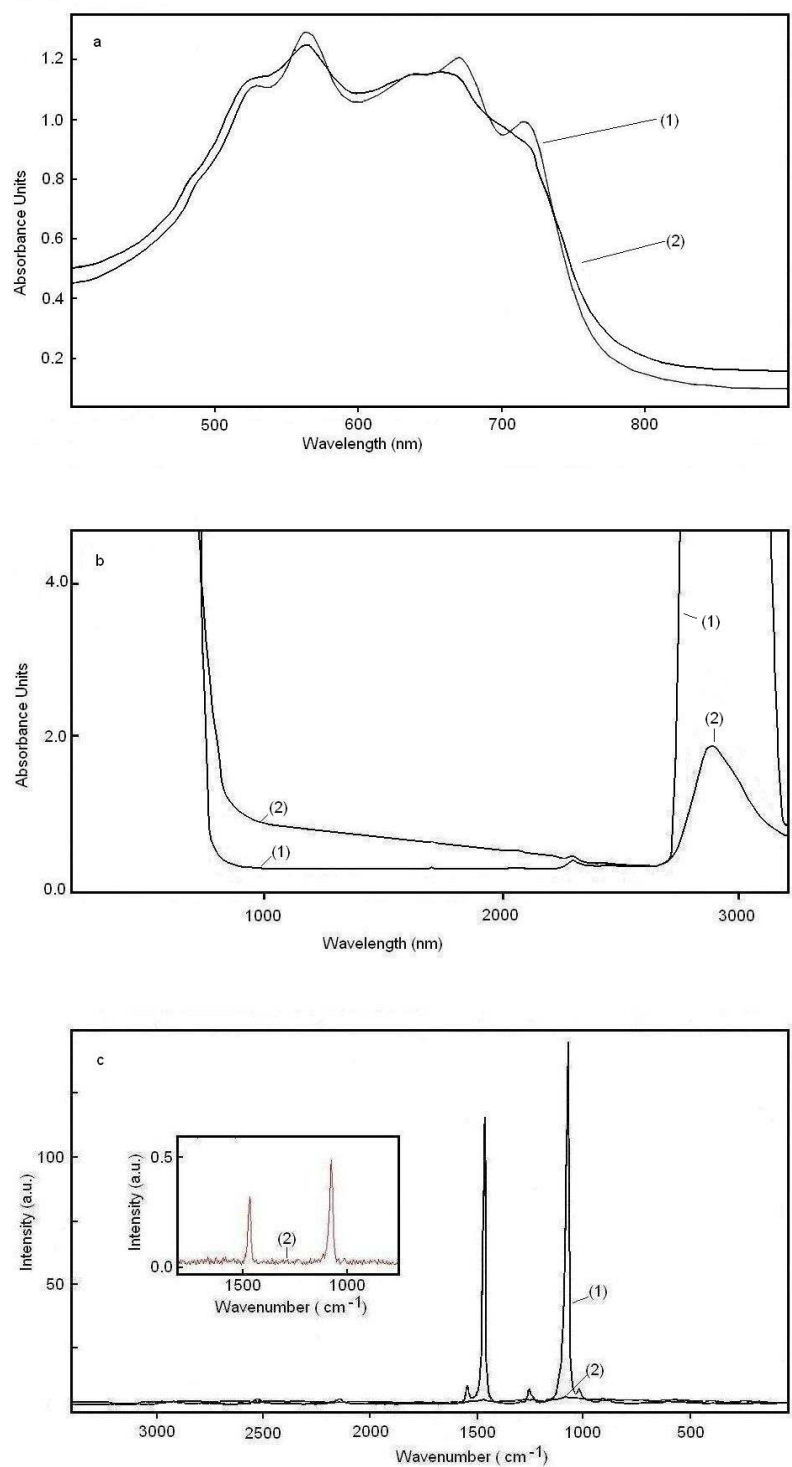

**Figure 1 | Optical characteristics of NPA films at the air (curve 1) and in vacuum (curve 2). 1a**-Electronic absorption spectra of 10 μm thickness film. **1b-**Near-IR spectra of 250 μm thickness film**. 1c-**Raman scattering spectra of 500 μm thickness film.

All three spectra display a growing intensity of background absorption, and a reduction of S/N ratio. Between these results and results published earlier [12] one can observe that spectral characteristics of NPA films after pumping out become similar to those of free-standing PA films. This is not surprising given that spectra of free-standing PA films almost always record in vacuum. However, the spectral characteristics of NPA films taken off on air (Fig.1, curves 1) are indeed highly surprising. Vibrational structure of absorption band of trans- PA (Fig.1a, curve 1) is observed, along with a transparency window in near-IR field (Fig.1b, curve 1), and an abnormally high intensity of Raman scattering (Fig.1c, curve 1). None of these traits have been otherwise observed in PA.

5. Raman scattering of NPA films.

Of the characteristics of NPA films under discussion, pumping out exerts the greatest influence on Raman scattering properties. Intensity of Raman scattering differs by three orders of magnitude between films freshly prepared and films pumped out. Additionally, pumping out leads to a sharp decrease in film stability under laser irradiation. Figure 2 displays Raman scattering spectra of a newly prepared NPA film recorded on air at beam powers between 20mW and 200 mW. One can see from the figure, that the increase in beam power is accompanied by linear growth in intensity of both bands of NPA fundamental vibrations. However, even tenfold growth in intensity is only accompanied by insignificant (<0,5 cm$^{-1}$) widening of the bands. It has been shown earlier [15] that the bands of fundamental vibrations in Raman scattering spectra of NPA are thermochromic, with coefficient of ~0,05 cm$^{-1}$/K. As such, prolonged (> 30 minutes) irradiation of films with optical density ($A_{1064nm}$ = 1) by a laser beam of 2,5 kW/cm$^2$ results in very insignificant (<10$^{o}$C) heating. Pumping out of NPA films leads to radical change in their interaction with the beam. Prolonged (> 10 minutes) irradiation by a 20 mW beam leads to significant widening of the bands, and decrease in intensity of Raman scattering. At powers >30 mW irradiation actually results in destruction of the films within 1 minute. Such a low stability of NPA films in vacuum does not afford to register the dependence of Raman scattering intensity on power of the exciting beam in conditions similar to those used on the air.

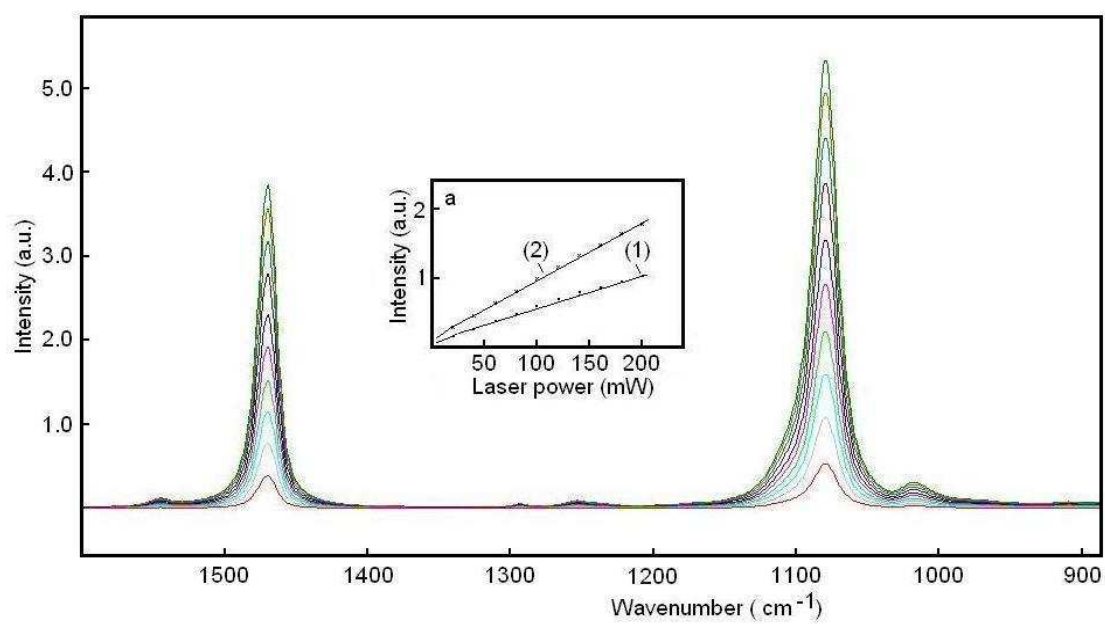

**Figure 2 | Raman scattering spectra of 500 μm thickness NPA film at power of excitation from 20 to 200 mW. 2a-**Dependence of intensity of fundamentals from excitation power. Vibrations of double bond (curve 1); single bond (curve 2).

6. IR spectra of NPA films. Conditions of water de-sorption.

Pumping out leads to de-sorption of water from NPA films accompanied by characteristic reduction of intensity of band ~2900 nm in near-IR spectrum (Fig.1b). Our research has shown that water de-sorption occurs not only at pumping out of films, but also at exposure to arid atmosphere. Figure 3 displays IR spectra of 500-μm thick NPA film, placed in cuvette chamber blown over by dry nitrogen. De-sorption takes about 1,5 hours for 500μm thick film, and is characterized by a significant change of structure of ~3400 cm$^{-1}$ band during reduction of its intensity. Figure 3a displays a fragment of IR spectrum of 50-μm thick NPA film record in atmosphere with relative humidity of 50% (curve 1), and after 1 hour of exposure to dry nitrogen (curve 2).

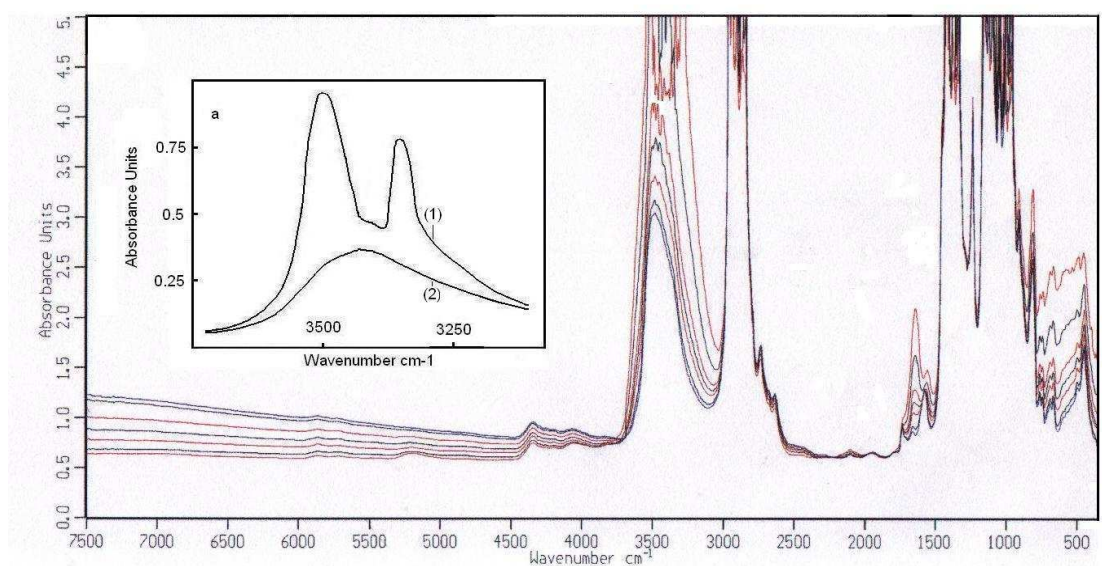

**Figure 3 | IR spectra of 500 μm thickness NPA film in arid atmosphere.** Time of spectrum registration – 1 minute. Time interval between spectra – 12 minutes. **3a**- Fragment of IR spectra of 50 μm thickness NPA film before (curve 1) and after (curve 2) 1 hour exposure in arid atmosphere.

One can see that in humid atmosphere the band at 3400 cm$^{-1}$ is observed as two well resolved maxima at 3504 cm$^{-1}$ and 3351 cm$^{-1}$. In the process of de-sorption (Fig.3), intensity of these maxima decreases as they are smoothly displaced to the centre of a bending around band. After 1.5 hours in dry nitrogen both maxima disappear; only the polyvinylbutyral band at 3431 cm$^{-1}$ is retained. The process is reversible; dehydrated NPA films absorbed almost the same quantity of water as the initial films. However, at repeated absorption of water, a band at ~3400 cm$^{-1}$ is observed as a wide maximum without fine structure. The optical characteristics of NPA films are not restored through repeated absorption.

We suggest that the initial NPA films contain water in two different forms [16,17]. One is adsorbed by polyvinylbutyral (band at 3504 cm$^{-1}$) and forms hydrogen bonds with its hydroxyl groups. The other (band at 3351 cm$^{-1}$) forms plasticized layers between polyacetylene nano-particles and the PVB matrix. It is also suggested that by repeated absorption water does not form such plasticized layers.

Results obtained let us assume that the unique optical characteristics of NPA films are connected with the 3D-ordering of nano-particles in the PVB matrix. Namely, the plasticized layers of water provide for rotational mobility of nano-particles required for synchronization of their interaction in the films.

## 7. Model.

Bulk polyacetylene contains conformational defects [18]. It is characterized by high electronic polarizability, presence of charged states in the band gap, and the ability to stabilize charged solitons at photo-excitation [19]. Basic intermolecular interactions in bulk polyacetylene are generated by Coulomb's interactions of charged defects, and fluctuations of charge density [20,21].

NPA does not contain conformational defects [12]. It possesses an abnormally high vibrational hyper-polarizability, characterized by the absence of charged states in the band gap, and does not form charged solitons at photo-excitation [22,23]. We assume that basic intermolecular interactions in NPA are generated by the interactions of coherently oscillated π-electrons of polyene chain with the oscillating electromagnetic field of the neighboring macromolecules.

Thermal or zero-point fluctuations excite modes of fundamental vibrations in macromolecules [24]. The fundamental vibrations in trans-NPA are coherent vibrations of single and double bonds [13]. Excitation of these modes is accompanied by ordered high frequency oscillations of π-electronic density of the polyene chain [14], which lead to appearance of oscillating electromagnetic field around macromolecules. This field, in turn, excites the π-electrons in the neighboring macromolecules, causing ordered oscillations of π-electronic density in them. Thus, each trans-polyene chain can be considered as a molecular conductor adjusted to the frequency of coherent oscillations. According to electromagnetic induction laws the energy of resonant interaction between parallel, coherently oscillating macromolecules should be proportional to frequency of oscillation squared to their vibrational hyper-polarizability, and inversely proportional to the square of distance. It should grow with increase in chain length and reduction of bond length alternation of macromolecules. Super high-frequency field oscillations are weakly shielded by organic media, and should be observed at significant distance between oscillating macromolecules. We shall term such interactions *coherent dispersive interactions* (CDI).

The above reasoning suggests that macromolecules of trans-PA in nano-particles are under the influence of two opposing forces. Peierl's instability [25] aspires to increase the bond length alternation, and CDI – to reduce it. The balance of these forces defines temperature dependence of bond length alternation in nano-particles.

Vibrational structure of dilute solutions of NPA was studied in [15]. It was shown that a decrease in solution temperature leads to a change in the vibrational structure of trans-PA in electronic absorption spectra, and an appearance of low-frequency shoulder near the band 1470 cm$^{-1}$ in Raman scattering spectra. These changes flow in the direction opposite of what is characteristic of individual polyenes [26] and testify that decreasing temperature of nano-particles is accompanied by reduction in bond length alternation. As such, we assume that energy of CDI in nano-particles exceeds the energy of Peierl's instability of trans-PA chains. CDI increases vibrational hyperpolarizability and intensity of Raman scattering of macromolecules. It was shown [27] that the normalized intensity of Raman scattering of NPA in solution of n-butanol exceeds the intensity of hydrogen bonds of solvent by ~8000 times.

During the formation of films the distance between nano-particles decreases, and they form a common, oscillating electromagnetic field. The stipulation for resonant interactions between nano-particles is the coincidence of frequencies of their oscillations [25,28], which depend on dielectric characteristics of the media [29]. Therefore any type of disorder [30] in the film should lead to deterioration of CDI, and consequently, of NPA's spectral characteristics. However, spectral characteristics of newly prepared NPA films on air surpass even the spectral characteristics of solutions [12]. We suppose, therefore, that the process of film formation is accompanied by coherent interaction and 3D ordering of nano-particles under the action of CD forces.

Dielectric characteristics of macromolecules are strongly anisotropic. Besides the change of bond length alternation of macromolecules should be accompanied by change of nano-particles volume. The necessary condition of CDI between nano-particles is the ability of media to support necessary structural changes. It is probable that clusters waters on the border of nano-partiles with the matrix provide the high vibrational and rotational mobility of nano-particles, and allow the manifestation of the unique optical properties of NPA.

# 8. References.

Asknowledgements: This work was supported by award from USA Air Force Research Laboratory, Materials & Manufacturing Directorate. I thank Douglas Dudis and his colleagues for help. I also thank D. Parschuk and A. Zuzin for helpful discussions.

Correspondence: Correspondence and requests for materials should be addressed to Valerii Kobryanskii (kobryan@juno.com).